\documentclass[a4paper,11pt]{article}

\usepackage{jcappub}
\usepackage{hyperref}
\usepackage{enumitem}
\usepackage{dcolumn}
\usepackage{bm}
\usepackage[latin1]{inputenc}
\usepackage[spanish,english]{babel}
\usepackage{amsfonts}
\usepackage{amssymb}
\usepackage{amsmath}
\usepackage{graphicx}
\usepackage{setspace}
\usepackage[T1]{fontenc} 

\newcommand{\be}{\begin{equation}}
\newcommand{\ee}{\end{equation}}
\newcommand{\bea}{\begin{eqnarray}}
\newcommand{\eea}{\end{eqnarray}}

\newcommand{\E}{\mathcal{E}}

\title{\boldmath  Semiclassical geons at particle accelerators}

\author[a]{Gonzalo J. Olmo}
\author[b]{D. Rubiera-Garcia}

\affiliation[a]{Departamento de F\'{i}sica Te\'{o}rica and
IFIC,  Universidad de Valencia-CSIC, Facultad de F\'{i}sica, C/ Dr. Moliner 50,
Burjassot-46100, Valencia, Spain.}
\affiliation[b]{Departamento de F\'{i}sica, Universidade Federal da
Para\'{i}ba, 58051-900 Jo\~ao Pessoa, Para\'{i}ba, Brazil}

\emailAdd{gonzalo.olmo@csic.es}
\emailAdd{drubiera@fisica.ufpb.br}

\abstract{We point out that in certain four-dimensional extensions of general relativity constructed within the Palatini formalism stable self-gravitating objects with a discrete mass and charge spectrum may exist. The incorporation of nonlinearities in the electromagnetic field may effectively reduce their mass spectrum by many orders of magnitude. As a consequence, these objects could be within (or near) the reach of current particle accelerators. We provide an exactly solvable model to support this idea.}

\keywords{Modified gravity, Palatini approach, nonsingular spacetimes, semiclassical geons, Born-Infeld, wormholes}

\arxivnumber{1306.6537 [hep-th]}

\begin{document}

\maketitle
\flushbottom

\section{Introduction}


Black holes are a genuine prediction of general relativity (GR) that set the classical limits of validity of Einstein's theory itself in the high-curvature regime defined by their  singularity. 
They also pose severe challenges to a consistent description of physical phenomena in which gravitation and quantum physics must be combined to give a reliable picture. Black holes are thus regarded as a means to explore new physics where gravitation approaches the quantum regime and to understand the fate of quantum information in the presence of event horizons and singularities. The existence of extra dimensions required by the string/M theory approach would also affect the dynamics of black hole formation/evaporation and the physics of elementary particles at very high energies. For these reasons, the physics of black holes and, in particular, of microscopic black holes is nowadays a very active field of research from both theoretical and experimental perspectives. In fact, it has been suggested that microscopic black holes could be created in particle accelerators \cite{BHaccelerators1,BHaccelerators2,BHaccelerators3} and numerous studies have been carried out to determine the feasibility of their production \cite{Rezzolla:2012nr,East:2012mb,Cavaglia:2002si,Yoshino:2002tx,Koch} and to characterize their observational signatures \cite{Khachatryan:2010wx,Cavaglia:2007ir}. In this sense, the general view is that
extra dimensions are required to reduce the effective Planck mass to the TeV energy scale \cite{TeV1,TeV2,TeV3,TeV4} (see also the review \cite{TeVreview}), which would make them accessible through current or future particle accelerators. Once produced, the semiclassical theory predicts that angular momentum and charge should be radiated away very rapidly yielding a Schwarzschild black hole as the late-time phase of the process. Since black holes are quantum mechanically unstable under Hawking decay, the Schwarzschild phase could be observed  through the detection of a burst of Hawking quanta.

The above picture is expected to be robust for sufficiently massive black holes, for which the backreaction on the geometry is weak and can be neglected. In this approach, one quantizes the matter fields on top of the classical geometry of GR and neglects the change in the geometry produced by the quantum evaporation process. It should be noted, in this sense, that the renormalizability of the matter fields in a curved background requires a completion of the gravity Lagrangian that involves quadratic corrections in the curvature tensor \cite{field-quantization1, field-quantization2}. Moreover, these corrections also arise in several approaches to quantum gravity, such as those based on string theory \cite{strings,Ortin}. On consistency grounds, the semiclassical description should thus take into account these corrections even in the case in which the matter backreaction is neglected. However, due to the fact that quadratic curvature corrections lead to higher-order derivative equations, which makes it difficult to find exact solutions and generates instabilities and causality violations, the standard GR description is generally preferred.

In this sense, we have recently investigated \cite{or12a,or12b,or12c} how the structure of black holes with electric charge gets modified when a quadratic extension of GR is formulated \`{a} la Palatini, i.e., without imposing any {\it a priori } relation between the metric and affine structures of space-time \cite{olmo11}.
This approach avoids important shortcomings of the usual formulation of the  quadratic theory, in which the connection is defined {\it a priori} as the Christoffel symbols of the metric. Relaxing the Levi-Civita condition on the connection, one generically finds the existence of invariant volumes associated to the connection, which define a metric structure algebraically related with that defined by the metric $g_{\mu\nu}$ \cite{or13}. As a result, the Palatini version yields second-order equations for $g_{\mu\nu}$, thus avoiding ghosts and other instabilities. In fact, in vacuum the field equations exactly boil down to those of GR, which guarantees that Minkowski is a stable vacuum solution and that there are no new propagating degrees of freedom beyond the typical spin-2 massless gravitons. The dynamics differs from that of GR in nonlinearities induced by the matter on the right-hand side of the equations. These nonlinear terms arise due to the nontrivial role played by the matter in the determination of the connection \cite{or13}.

In our analysis \cite{lor12}, we found that spherically symmetric, electro-vacuum solutions can be naturally interpreted as geons, i.e., as consistent solutions of the gravitational-electromagnetic system of equations without sources. This is possible thanks to the nontrivial topology of the resulting space-time, which through the formation of a wormhole allows to define electric charges without requiring the explicit existence of sources of the electric field. In this scenario, massive black holes are almost identical in their macroscopic properties to those found in GR. However, new relevant structures arise in the lowest band of the mass and charge spectrum (microscopic regime). In particular, below a certain critical charge $q_c=e N_c$, with $N_c= \sqrt{2/\alpha_{em}}\approx 16.55$, where $\alpha_{em}$ is the fine structure constant and $e$ the electron charge, one finds a set of solutions with no event horizon and with smooth curvature invariants everywhere.  Moreover, the mass of these solutions can be exactly identified with the energy stored in the electric field and their action (evaluated on the solutions) coincides with that of a massive point-like particle at rest.  The topological character of their charge, therefore, makes these solutions stable against arbitrary perturbations of the metric as long as the topology does not change. On the other hand, the absence of an event horizon makes these configurations stable against Hawking decay (regular solutions with an event horizon also exist, though they are unstable). These properties together with the fact that these solutions lie in the lowest band of the charge and mass spectrum of the theory suggest that they can be naturally identified as black hole remnants. The existence of such solutions in a minimal extension of GR demands further research to better understand their stability and chances of being experimentally accessible. This is the main motivation for this work.

In this paper we study the effects that modifications on the matter sector of the theory studied in \cite{lor12} could have for the qualitative and quantitative stability of its solutions. In \cite{lor12}, we considered the quadratic Palatini theory
\begin{equation}\label{eq:gravtheory}
S_g=\frac{c^3}{16\pi G}\int d^4x \sqrt{-g}\left[R + l_P^2(a R^2+ b R_{\mu\nu}R^{\mu\nu})\right]
\end{equation}
where $l_P^2=\hbar G/c^3$, coupled to a spherically symmetric, sourceless electric field. The quadratic curvature terms are regarded as quantum gravitational corrections\footnote{Other types of quantum corrections in the gravitational sector in scenarios inspired by loop quantum gravity have been considered in \cite{Modesto1, Modesto2, Modesto3}.} that vanish when $\hbar\to 0$. In that limit the theory recovers the usual Einstein-Maxwell equations, which have the Reissner-Nordstr\"om black hole as a solution. For finite $\hbar$ new solutions arise with the properties summarized above. Though the Reissner-Nordstr\"om solution is generally accepted as a valid solution of the classical Einstein-Maxwell system, the fact is that in the innermost regions of such black holes the amplitude of the electric field grows without bound above the threshold of quantum pair production \cite{Schwinger51}, which should have an impact on the effective description of the electric field. These effects were neglected in  \cite{lor12}. Here we want to explore the consequences for the existence and properties of the Palatini-Maxwell geons of \cite{lor12} when nonlinearities in the description of the electromagnetic field are incorporated.

It is well known that under certain conditions \cite{Dunne04} the effects of the quantum vacuum can be taken into account within the effective Lagrangians approach. In this approach, once the heavy degrees of freedom are integrated out in the path integral of the original action of quantum electrodynamics, a perturbative expansion leads to series of effective (classical) Lagrangians correcting the Maxwell one, which take the form of powers in the field invariants of the electromagnetic field \cite{Dobado}. This is the case of the Euler-Heisenberg Lagrangian \cite{EH1,EH2}. These effective Lagrangians account, at a purely classical level, not only for the dynamics of the low-energy fields, but also for the quantum interactions with the removed heavy-mode sector. Historically, the nonlinear modifications on the dynamics of the electromagnetic field date back to the introduction of the Born-Infeld Lagrangian \cite{BI34}, aimed at the removal of the divergence of the electron's self-energy in classical electrodynamics. It has been shown that this Lagrangian also arises in the low-energy regime of string and D-Brane physics \cite{string-NEDs1,string-NEDs2,string-NEDs3,string-NEDs4,string-NEDs5}, and it is often considered in the context of black hole physics in GR \cite{BI-GR1,BI-GR2,BI-GR3} and in modified theories of gravity \cite{BI-modifiedgravity1,BI-modifiedgravity2,BI-modifiedgravity3}. For concreteness and analytical simplicity, in this work we will consider the gravity theory (\ref{eq:gravtheory}) coupled to the Born-Infeld nonlinear theory of electrodynamics as a way to test the robustness of our previous results under quantum-motivated modifications of the matter sector.

The Born-Infeld theory recovers the linear Maxwell theory when a certain parameter $\beta^2$ is taken to infinity (see section \ref{sec:VI} below for details). For any finite value of $\beta^2$ the nonlinearities of the matter sector enter in the construction of the geometry.  As a result, here we will see that the qualitative features of the solutions found in  \cite{lor12} persist for arbitrary values of $\beta^2$. However, from a quantitative point of view, we find that the mass of the remnants can be many orders of magnitude smaller than that found in the Maxwell case, while the maximum charge allowed to have a remnant can be much larger. These results, therefore, put forward that the Maxwell case is the least favorable scenario from an experimental perspective and that quantum corrections in the matter sector can substantially improve the chances of experimentally detecting the kind of black hole remnants found in \cite{lor12}. In fact, for specific values of the parameter $\beta^2$, we find that the mass spectrum of the stable solutions can be lowered from the Planck scale, $\sim 10^{19}$ GeV, down to the TeV scale. This means that ongoing experiments in particle accelerators such as the LHC could be used to explore new gravitational phenomena directly related with the Planck scale within a purely four-dimensional scenario.

The paper is organized as follows: in Sec. \ref{sec:II} we define our theory and  provide the general metric and connection field equations. We introduce the matter sector of our theory under the form of a nonlinear electromagnetic field in Sec. \ref{sec:III}, and specify the Palatini field equations for this matter in Sec. \ref{sec:IV}. These equations are solved in Sec. \ref{sec:V} for a spherically symmetric metric. In Sec. \ref{sec:VI} we introduce the Born-Infeld theory. In Sec. \ref{sec:VII} we formulate the metric components for the quadratic Palatini theory and study an exactly solvable case in Sec. \ref{sec:VIII}. We conclude in Sec. \ref{sec:IX} with some final remarks.

\section{Basics of Palatini gravity} \label{sec:II}

Let us consider a general family of Palatini theories defined as
\begin{equation}\label{eq:action}
S[g,\Gamma,\psi_m]=\frac{1}{2\kappa^2}\int d^4x \sqrt{-g}f(R,Q) +S_m[g,\psi_m],
\end{equation}
where $f(R,Q)$ represents the gravity Lagrangian, $\kappa^2$ is a constant with suitable dimensions (in GR, $\kappa^2 \equiv 8\pi G$),  $S_m[g,\psi_m]$ represents the matter action with $\psi_m$ representing the matter fields, $g_{\alpha\beta}$ is the space-time metric,  $R=g^{\mu\nu}R_{\mu\nu}$, $Q=g^{\mu\alpha}g^{\nu\beta}R_{\mu\nu}R_{\alpha\beta}$, $R_{\mu\nu}={R^\rho}_{\mu\rho\nu}$,   and
\begin{equation}\label{eq:Riemann}
{R^\alpha}_{\beta\mu\nu}=\partial_{\mu}
\Gamma^{\alpha}_{\nu\beta}-\partial_{\nu}
\Gamma^{\alpha}_{\mu\beta}+\Gamma^{\alpha}_{\mu\lambda}\Gamma^{\lambda}_{\nu\beta}-\Gamma^{\alpha}_{\nu\lambda}\Gamma^{\lambda}_{\mu\beta} \ .
\end{equation}
In the above action (\ref{eq:action}) the independent connection $\Gamma_{\mu\nu}^{\lambda}$ is, for simplicity, not directly coupled to the matter sector $S_m$. In the electromagnetic case to be discussed in this paper this point is irrelevant if the connection is symmetric but in more general cases the coupling must be specified. We also assume vanishing torsion, i.e., $\Gamma^\lambda_{[\mu\nu]}=0$, though impose this condition {\it a posteriori}, once the field equations have been obtained (see \cite{or13} for details). This implies that the Ricci tensor is symmetric, i.e., $R_{[\mu\nu]}=0$. Thus, in what follows symmetry in the indices of $R_{\mu\nu}$ will be implicitly understood. It must be stressed that the Palatini formulation of (\ref{eq:action}) is inequivalent to the metric approach, which translates into a different structure of the field equations \cite{Borunda08,Borunda08b} and, consequently, on the mathematical and physical features of the theory.

Under the above assumptions, independent variations of the action (\ref{eq:action}) with respect to metric and connection yield
\begin{eqnarray}
f_R R_{\mu\nu}-\frac{f}{2}g_{\mu\nu}+2f_QR_{\mu\alpha}{R^\alpha}_\nu &=& \kappa^2 T_{\mu\nu}\label{eq:met-varX}\\
\nabla_{\beta}\left[\sqrt{-g}\left(f_R g^{\mu\nu}+2f_Q R^{\mu\nu}\right)\right]&=&0  \ ,
 \label{eq:con-varX}
\end{eqnarray}
where we have used the short-hand notation $f_R \equiv \frac{df}{dR}$ and $f_Q \equiv \frac{df}{dQ}$. As a first step to solve Eq.(\ref{eq:met-varX}) we introduce the matrix $\hat{P}$, whose components are ${P_\mu}^\nu\equiv R_{\mu\alpha}g^{\alpha\nu}$, which allows us to express (\ref{eq:met-varX}) as
\begin{equation}
f_R {P_\mu}^\nu-\frac{f}{2}{\delta_\mu}^\nu+2f_Q{P_\mu}^\alpha {P_\alpha}^\nu= \kappa^2 {T_\mu}^\nu\label{eq:met-varRQ1} \ .
\end{equation}
In matrix notation, this equation reads
\begin{equation}
2f_Q\hat{P}^2+f_R \hat{P}-\frac{f}{2}\hat{I} = \kappa^2 \hat{T} \label{eq:met-varRQ2} \ ,
\end{equation}
where $\hat{T}$ is the matrix representation of ${T_\mu}^\nu$. Note also that in this notation we have $R={[\hat{P}]_\mu}^\mu$ and $Q={[\hat{P}^2]_\mu}^\mu$. Therefore (\ref{eq:met-varRQ2}) can be seen as a nonlinear algebraic equation for the object $\hat{P}$ as a function of the energy-momentum tensor, i.e., $\hat{P}=\hat{P}(\hat{T})$. We will assume that a valid solution to that equation can be found. Expressing (\ref{eq:con-varX}) as
\begin{equation}
\nabla_{\beta}\left[\sqrt{-g}g^{\mu\alpha}\left(f_R \delta_\alpha^\nu+2f_Q {P_\alpha}^{\nu}\right)\right]=0 \ ,
\end{equation}
we see that the term in brackets only depends on the metric $g_{\mu\nu}$ and  $\hat T$, which implies that the connection $\Gamma_{\mu\nu}^{\lambda}$ appears linearly in this equation and can be solved by algebraic means. Defining now the object
\begin{equation} \label{eq:geometrymissmatch}
{\Sigma_\alpha}^{\nu}=\left(f_R \delta_{\alpha}^{\nu} +2f_Q {P_\alpha}^{\nu}\right),
\end{equation}
we can rewrite the connection equation in the form
\begin{equation} \label{eq:auxmetric}
\nabla_{\beta}[\sqrt{-h} h^{\mu\nu}]=0,
\end{equation}
where
\begin{equation} \label{eq:metricsrelation}
\sqrt{h} {h}^{\mu\nu}= \sqrt{-g} {g}^{\mu\alpha} {\Sigma_\alpha}^\nu \ .
\end{equation}
Eq.(\ref{eq:auxmetric}) implies that $\Gamma^\alpha_{\mu\nu}$ is the Levi-Civita connection of $h_{\mu\nu}$.  From the relation (\ref{eq:metricsrelation}) one finds that $h_{\mu\nu}$ and $g_{\mu\nu}$ are related by
\begin{equation} \label{eq:h-g}
h^{\mu\nu}=\frac{g^{\mu\alpha}{\Sigma_{\alpha}}^\nu}{\sqrt{\det \hat{\Sigma}}} \ , \quad
h_{\mu\nu}=\left(\sqrt{\det \hat{\Sigma}}\right){\Sigma_{\mu}}^{\alpha}g_{\alpha\nu} \ .
\end{equation}
This puts forward that ${\Sigma_\mu}^\nu$ defines the relative deformation existing between the physical metric $g_{\mu\nu}$ and the auxiliary metric $h_{\mu\nu}$.

Using the definition of ${\Sigma_\mu}^\nu$ and the relations (\ref{eq:h-g}), it is easy to see that (\ref{eq:met-varX}) [or, alternatively, (\ref{eq:met-varRQ1})] can be written as ${P_\mu}^\alpha {\Sigma_\alpha}^\nu=R_{\mu\alpha}h^{\alpha\nu} \sqrt{\det \hat\Sigma}=\frac{f}{2}{\delta_\mu^\nu}+{T_\mu}^\nu$, which allows to express the metric field equations using $h_{\mu\nu}$ as follows
\begin{equation} \label{eq:fieldequations}
{R_{\mu}}^{\nu}(h)=\frac{1}{\sqrt{\det \hat{\Sigma}}}\left(\frac{f}{2}{\delta_{\mu}}^{\nu}+ \kappa^2 {T_{\mu}}^{\nu} \right) \ .
\end{equation}
Note that since $R, Q$, and $ \hat{\Sigma}$ are functions of $\hat T$, then the right-hand side of (\ref{eq:fieldequations}) is completely specified by the matter, whereas the left-hand side simply represents the Ricci tensor of the metric $h_{\mu\nu}$. The equations satisfied by $h_{\mu\nu}$, therefore, are formally very similar to those found in GR. We emphasize that since $h_{\mu\nu}$ satisfies second-order equations, the algebraic relation existing between $h_{\mu\nu}$ and $g_{\mu\nu}$ gurantees that $g_{\mu\nu}$ also satisfies second-order equations. Note also that in vacuum  (\ref{eq:fieldequations}) boils down to GR plus an effective cosmological constant (see \cite{or13} for details). As in previous works \cite{lor12,or12a}, here we shall take the strategy of solving the field equations  in terms of $h_{\mu\nu}$ and then use $\hat{\Sigma}$ in (\ref{eq:h-g}) to obtain the physical metric $g_{\mu\nu}$.

\section{The matter sector} \label{sec:III}

For the sake of generality,  we define the matter sector of our theory by means of the action
\begin{equation} \label{eq:matter}
S_m=\frac{1}{8\pi} \int d^4x \sqrt{-g} \varphi(X,Y),
\end{equation}
where $\varphi(X,Y)$ represents a (so far unspecified) Lagrangian of the two field invariants $X=-\frac{1}{2}F_{\mu\nu}F^{\mu\nu}$ and $Y=-\frac{1}{2}F_{\mu\nu}{*F}^{\mu\nu}$ that can be constructed with the field strength tensor $F_{\mu\nu}=\partial_{\mu}A_{\nu}-\partial_{\nu}A_{\mu}$ of the vector potential $A_{\mu}$, and its dual $*F^{\mu\nu}=\frac{1}{2}\epsilon^{\mu\nu\alpha\beta}F_{\alpha\beta}$. By now this function is only constrained by parity invariance ($\varphi(X,Y)=\varphi(X,-Y)$). Let us stress that (\ref{eq:matter}) is the natural generalization of the Maxwell action
\begin{equation} \label{eq:action-matter}
S_m=-\frac{1}{16\pi} \int d^4x \sqrt{-g} F_{\alpha\beta}F^{\alpha\beta},
\end{equation}
to the general charged case. When $\varphi(X,Y)\neq X$, one says that (\ref{eq:matter}) defines a nonlinear electrodynamics (NED) theory.
The energy-momentum tensor derived from (\ref{eq:matter}) and appearing in Eq.(\ref{eq:fieldequations}) is written as
\begin{equation} \label{eq:em1}
{T_{\mu}}^{\nu}=-\frac{1}{4\pi}\left[\varphi_X {F_{\mu}}^{\alpha}{F_{\alpha}}^{\nu}+ \varphi_Y {F_{\mu}}^{\alpha}{*F_{\alpha}}^{\nu}-\frac{\delta_{\mu}^{\nu}}{4}\varphi(X,Y)\right].
\end{equation}
On the other hand, the field equations for this NED matter are written as
\begin{equation}
\nabla_{\mu} \left( \sqrt{-g} (\varphi_X F^{\mu\nu} + \varphi_Y {*F}^{\mu\nu} \right))=0.
\end{equation}
In what follows we shall only deal with purely electrostatic configurations, for which the only nonvanishing component of the field strength tensor is $F^{tr}(r)$. This ansatz implies that the $Y$-invariant will play no role in the dynamics of the theory and can be safely neglected from now on. In such a case, assuming a line element of the form $ds^2=g_{tt}dt^2+g_{rr}dr^2+r^2d\Omega^2$ the field equations lead to
\begin{equation} \label{eq:first-integral}
F^{tr}=\frac{q}{r^2 \varphi_X \sqrt{-g_{tt}g_{rr}}},
\end{equation}
where $q$ is an integration constant interpreted as the electric charge associated to a given solution. Writing $X=-g_{tt}g_{rr} (F^{tr})^2$ it follows that for any spherically symmetric metric
\begin{equation}
\varphi_X^2 X=\frac{q^2}{r^4}.
\end{equation}
On the other hand, the components of the energy-momentum tensor (\ref{eq:em1}) for these electrostatic solutions read, in matrix form
\begin{equation}\label{eq:Tmn-EM}
{T_\mu}^\nu=\frac{1}{8\pi}\begin{pmatrix}
 (\varphi-2X\varphi_X) \hat{I} & \hat{0}  \\
\hat{0} & \varphi \hat{I}
\end{pmatrix},
\end{equation}
where $\hat{I}$ and $\hat 0$ are the identity and zero $2 \times 2$ matrices, respectively.

In order to write the field equations associated to the matter source (\ref{eq:matter}), we first need to find the explicit form of $\hat{P}$ for it, which will allow us to construct $\hat{\Sigma}$ and compute its determinant. To do this, we write (\ref{eq:met-varRQ2}) as

\begin{equation}
2f_Q\left(\hat{P}+\frac{f_R}{4f_Q}\hat{I}\right)^2=
\begin{pmatrix}
\lambda_-^2\hat{I} & \hat{0}  \\
\hat{0} & \lambda_+^2\hat{I}
\end{pmatrix}  \ ,
\end{equation}
where

\begin{eqnarray}
\lambda_{+}^2&=&\frac{1}{2}\left(f+\frac{f_R^2}{4f_Q}+2k^2 T_{\theta}^{\theta}\right)= \frac{1}{2}\left(f+\frac{f_R^2}{4f_Q}+\tilde{\kappa}^2\varphi \right) \label{lambda+}\\
\lambda_{-}^2&=&\frac{1}{2} \left(f+\frac{f_R^2}{4f_Q} +2k^2 T_{t}^{t}\right)= \frac{1}{2} \left(f+\frac{f_R^2}{4f_Q} +\tilde{\kappa}^2\left(\varphi-2X\varphi_X\right)\right),\label{lambda-}
\end{eqnarray}
and we have defined $\tilde{\kappa}^2=\kappa^2/4\pi$. There are $16$ square roots that satisfy this equation, namely,
\begin{equation}
\sqrt{2f_Q}\left(\hat{P}+\frac{f_R}{4f_Q}\hat{I}\right)=
\begin{pmatrix}
s_1\lambda_- & 0 & 0 & 0  \\
0 & s_2 \lambda_- & 0 & 0\\
0 & 0 & s_3\lambda_+ & 0 \\
0 & 0 & 0 & s_4\lambda_+
\end{pmatrix}  \ ,
\end{equation}
where $s_i=\pm 1$. Agreement with GR in the low curvature regime (where $f_R\to 1$ and $f_Q\to 0$)  requires $s_i=+1$. For this reason, we simplify the notation and take
\begin{equation}\label{eq:M_ab}
\sqrt{2f_Q}\left(\hat{P}+\frac{f_R}{4f_Q}\hat{I}\right)=
\begin{pmatrix}
\lambda_- \hat{I}& \hat{0} \\
\hat{0} & \lambda_+ \hat{I}
\end{pmatrix}  \ .
\end{equation}
From this it follows that the matrix $\hat{\Sigma}$ is given by
\begin{equation} \label{eq:sigma-matrix}
\hat{\Sigma}=\frac{f_R}{2}\hat{I}+\sqrt{2f_Q}
\begin{pmatrix}
\lambda_- \hat{I}& \hat{0} \\
\hat{0} & \lambda_+ \hat{I}
\end{pmatrix}=\begin{pmatrix}
\sigma_- \hat{I}& \hat{0} \\
\hat{0} & \sigma_+\hat{I}
\end{pmatrix} \ ,
\end{equation}
where $\sigma_\pm=\left(\frac{f_R}{2}+\sqrt{2f_Q}\lambda_\pm\right)$ and $\lambda_{\pm}$ are given in Eqs.(\ref{lambda+}) and (\ref{lambda-}).

\section{Palatini gravity with matter} \label{sec:IV}

The field equations in matrix form for the energy-momentum tensor (\ref{eq:Tmn-EM}) are written as
\begin{equation}\label{eq:Rmn}
{R_{\mu}}^{\nu}(h)=\frac{1}{2 \sigma_{+} \sigma_{-}} \left(
\begin{array}{cc}
(f+2\kappa^2 T_t^t)\hat{I} & \hat{0} \\
\hat{0} & (f+2\kappa^2 T_{\theta}^{\theta}) \hat{I} \\
\end{array}
\right).
\end{equation}
In order to solve them we must compute explicitly the objects $\lambda_{\pm}^2$ and $\sigma_{\pm}$, besides $R$ and $Q$. To do this, we consider the quadratic Lagrangian
\begin{equation} \label{eq:gravitytheory}
f(R,Q)=R + l_P^2(a R^2+ b Q) \ ,
\end{equation}
and trace Eq.(\ref{eq:met-varX}) with $g_{\mu\nu}$, which yields
\begin{equation} \label{eq:curvaturescalar}
R=-\kappa^2 T,
\end{equation}
where $T$ is the trace of the energy-momentum tensor of the matter. In the Maxwell case, the $R-$dependent part of the gravity Lagrangian does not play a very relevant role in the dynamics as a consequence of the tracelessness of the energy-momentum tensor entering in Eq.(\ref{eq:curvaturescalar}). However, for NED the trace reads
\begin{equation} \label{eq:trace}
T=2T_t^t+ 2T_{\theta}^{\theta} =\frac{1}{2\pi}[\varphi-X\varphi_X] \ ,
\end{equation}
which is nonvanishing for any nonlinear function $\varphi(X,Y=0)$. Therefore NED make possible to excite new dynamics associated to the $R-$dependent part of the Lagrangian. The explicit expression for $Q$ can be obtained for the theory (\ref{eq:gravitytheory}) by taking the trace in (\ref{eq:M_ab}) and solving for $Q$, which leads to the result

\begin{equation} \label{eq:Q-invariant}
\hat{Q}= \frac{T^2}{4}+\frac{(T_t^t- T_{\theta}^{\theta})^2}{(1-(2a+b)\chi T)^2},
\end{equation}
where we have defined $\hat{Q}=Q/\kappa^4$ and introduced the parameter $\chi=\kappa^2 l_P^2 $. Next we can write $\lambda_{\pm}$ in Eqs.(\ref{lambda+}) and (\ref{lambda-}) as
\begin{eqnarray} \label{eq:lambda}
\lambda_{+}&=& \frac{1}{2\sqrt{2b}l_P}\Big(1-4\chi\frac{(aT^{\theta}_{\theta}+(b+a) T_{t}^{t})- (b+2a)^2 \chi (T_t^t+T_{\theta}^{\theta})^2}{1-2\chi (b+2a)(T_t^t+T_{\theta}^{\theta})} \Big) \\
\lambda_{-}&=& \lambda_{+}(T_t^t \leftrightarrow T_{\theta}^{\theta}),
\end{eqnarray}
and therefore the factors $\sigma_{\pm}$ appearing in (\ref{eq:sigma-matrix}) are obtained as

\begin{eqnarray} \label{eq:sigma1}
\sigma_{+}&=&1+2\chi\left[\frac{-[aT+bT^t_t]+\frac{(b+2a)(b+4a)}{4}\chi T^2}{1-(b+2a)\chi T} \right] \\
\sigma_{-}&=& \sigma_{+} (T_t^t \leftrightarrow T_{\theta}^{\theta}).\label{eq:sigma2}
\end{eqnarray}
When $b=0$ we obtain $\sigma_{\pm}=1-2a \chi \hat{T}$ which corresponds to the theory $f(R)=R+al_P^2 R^2$ coupled to NED charged matter, a case studied in \cite{fR-NED}. When $a=1$, $b \neq 0$ and vanishing trace ($T=0$, corresponding to Maxwell theory) we obtain $\sigma_+=1+ \chi \hat{T}_t^t$ and since $T_t^t=-T_{\theta}^{\theta}=-\frac{q^2}{8\pi r^4}$, defining a variable $z^4=\frac{4\pi}{\kappa^2l_P^2 \beta^2}r^4$ we achieve the result $\sigma_+=1+1/z^4$ and $Q=\frac{\tilde{\kappa}^4 q^4}{r^4}$, in agreement with the result obtained in \cite{or12a} for the Maxwell case.

\section{Solving the equations} \label{sec:V}

\subsection{The metric ansatz}

We shall first formally solve the field equations (\ref{eq:fieldequations}), which are valid for any $f(R,Q)$ gravity theory and any $\varphi(X,Y=0)$ matter Lagrangian densities. Later we will consider the particular case of Lagrangian (\ref{eq:gravitytheory}). For this purpose, it is convenient to introduce two different line elements in Schwarzschild-like coordinates, one associated to the physical metric $g_{\mu\nu}$,
\begin{equation}\label{eq:ds2g}
ds^2= g_{tt}dt^2+g_{rr}dr^2+r^2 d\Omega^2 \ ,
\end{equation}
and another associated to the auxiliary metric $h_{\mu\nu}$
\begin{equation}\label{eq:ds2h}
d\tilde{s}^2= h_{tt}dt^2+h_{rr}dr^2+\tilde{r}^2 d\Omega^2
\end{equation}
with $d\Omega^2= d\theta^2+\sin^2 (\theta) d \phi^2$. The relation between these two line elements is nontrivial due to the relation $g_{\mu\nu}=\Sigma_{\mu}^{\alpha}h_{\alpha\nu}/\sqrt{\det \Sigma}$. To be precise, they are related as

\begin{equation}\label{eq:g-vs-h}
g_{\mu\nu}=\begin{pmatrix}
g_{tt} & 0 & 0 & 0 \\
0 & g_{rr} & 0 & 0 \\
0 & 0 & r^2 & 0 \\
0 & 0 & 0 & r^2\sin^2 \theta
\end{pmatrix} = \begin{pmatrix}
\frac{h_{tt}}{\sigma_+} & 0 & 0 & 0 \\
0 & \frac{h_{rr}}{\sigma_+} & 0 & 0 \\
0 & 0 & \frac{h_{\theta\theta}}{\sigma_-} & 0 \\
0 & 0 & 0 & \frac{h_{\phi\phi}}{\sigma_-}
\end{pmatrix} \
.\end{equation}

From (\ref{eq:g-vs-h}), it is clear that $g_{tt}=h_{tt}/\sigma_+$, $g_{rr}=h_{rr}/\sigma_+$, and that $r^2$ and $\tilde{r}^2$ are related by $\tilde{r}^2 =r^2 \sigma_{-}$. We are particularly interested in those cases in which the function $\sigma_-$ may vanish at some $r=r_c>0$. When this happens, we find a center in the geometry defined by $h_{\mu\nu}$ (with $\tilde{r}=0$) and a two-sphere of area $A=4\pi r_c^2$ in the physical geometry. As we will see in further detail latter, the  physical geometry has a minimum at $r=r_c$ and, therefore, that surface can be identified as the throat of a wormhole. This can be easily seen by introducing a new coordinate $dx^2=dr^2 \sigma_{-}$ (with $x\in ]-\infty,\infty[$). The vanishing of $\sigma_{-}$ at $r_c$ is simply interpreted as the point where the radial function $r^2(x)$ reaches a minimum ($dr/dx=\sigma_-^{1/2}\to 0$) [see Fig.\ref{fig:x(z)} below]. 

\subsection{General solution}

Since the field equations take a simpler form in terms of the metric $h_{\mu\nu}$, see (\ref{eq:Rmn}), we will use the line element  (\ref{eq:ds2h})  to solve for the metric. For this purpose, it is convenient to  use $\tilde{r}$ as the radial coordinate, which brings  (\ref{eq:ds2h}) into
\begin{equation}\label{eq:ds2hbar}
d\tilde{s}^2= h_{tt}dt^2+h_{\tilde{r}\tilde{r}}d\tilde{r}^2+\tilde{r}^2 d\Omega^2 \ ,
\end{equation}
where $h_{rr}dr^2=h_{\tilde{r}\tilde{r}}d\tilde{r}^2$. Considering the redefinitions
$h_{tt}=-A(\tilde{r})e^{2\psi(\tilde{r})}, h_{\tilde{r}\tilde{r}}=1/A(\tilde{r})$, the components of the tensor ${R_{\mu}}^{\nu}(h)$ become
\begin{eqnarray}\label{eq:Rtt}
{R_t}^t&=&-\frac{1}{2h_{\tilde{r}\tilde{r}}}\Big[\frac{A_{\tilde{r}\tilde{r}}}{A}-\left(\frac{A_{\tilde{r}}}{A}\right)^2+
2\psi_{\tilde{r}\tilde{r}}+ \left(\frac{A_{\tilde{r}}}{A}+ 2\psi_{\tilde{r}}\right)\left(\frac{A_{\tilde{r}}}{A}+
\psi_{\tilde{r}}+\frac{2}{\tilde{r}}\right)\Big]\\
\label{eq:Rrr}
{R_{\tilde{r}}}^{\tilde{r}}&=&-\frac{1}{2h_{\tilde{r}\tilde{r}}}\Big[\frac{A_{\tilde{r}\tilde{r}}}{A}-
\left(\frac{A_{\tilde{r}}}{A}\right)^2+2\psi_{\tilde{r}\tilde{r}}+ \left(\frac{A_{\tilde{r}}}{A}+
2\psi_{\tilde{r}}\right)\left(\frac{A_{\tilde{r}}}{A}+\psi_{\tilde{r}}\right)+
\frac{2}{\tilde{r}}\frac{A_{\tilde{r}}}{A}\Big]\\
\label{eq:Rzz}
{R_\theta}^\theta&=&\frac{1}{\tilde{r}^2}\left[1-A(1+\tilde{r}\psi_{\tilde{r}})-\tilde{r}A_{\tilde{r}}\right] \ .
\end{eqnarray}
Now taking into account the symmetry ${T_t}^t={T_r}^r$ of the NED energy-momentum tensor, that holds for electrostatic spherically symmetric solutions, it is easily seen that the substraction ${R_t}^t-{R_{\tilde{r}}}^{\tilde{r}}$ in the field equations (\ref{eq:Rtt}) and (\ref{eq:Rrr}) leads to $\psi= constant$ and can be set to zero through a redefinition of the time coordinate, like in GR. This leaves a single equation to be solved, namely
\begin{equation} \label{eq:eomunique}
\frac{1}{\tilde{r}^2} \left(1-A(\tilde{r})-\tilde{r}A_{\tilde{r}} \right)=\frac{1}{2\sigma_+ \sigma_-}\left(f+\frac{\kappa^2}{4\pi} \varphi \right) \ .
\end{equation}
Taking the usual ansatz
\begin{equation}
A(\tilde{r})=1-\frac{2M(\tilde{r})}{\tilde{r}},
\end{equation}
and inserting it into (\ref{eq:eomunique}) we are led to

\begin{equation}
M_{\tilde{r}}=\frac{\tilde{r}^2}{4\sigma_{+}\sigma_{-}}\left(f+\frac{\kappa^2}{4\pi}\varphi \right).
\end{equation}
Using the relation between coordinates $\tilde{r}^2=r^2 \sigma_{-}$ we have $\frac{d\tilde{r}}{dr}=\sigma_{-}^{1/2}\left(1+\frac{r\sigma_{-,r}}{2\sigma_{-}} \right)$, and with this we collect the final result

\begin{equation} \label{eq:mass-r}
M_r=\frac{\left(f+\frac{\kappa^2}{4\pi}\varphi\right)r^2 \sigma_{-}^{1/2}}{4\sigma_+}\left(1+\frac{r\sigma_{-,r}}{2\sigma_-}\right).
\end{equation}
Let us recall that in the above derivation only two assumptions have been made, namely, spherical symmetry of the space-time and symmetry of the energy-momentum tensor of the matter ${T_t}^t={T_{r}}^{r}$, and thus are valid for any $f(R,Q)$ theory coupled to arbitrary charged NED matter $\varphi(X,Y=0)$. Moreover, Eq.(\ref{eq:mass-r}) completely determines the associated spherically symmetric, static solutions through a single function $M(r)$  once the gravity and matter Lagrangians are given.

Upon integration of the mass function (\ref{eq:mass-r}), one finds an expression of the form  $M(r)=M_0 + \Delta M$, where $M_0$ is an integration constant identified as the Schwarzschild mass $M_0\equiv r_S/2$, and $\Delta M$ represents the electromagnetic contribution. In order to deal with dimensionless magnitudes, we find it useful to introduce a function $G(r;a,b;\ldots)$ as follows
\begin{equation} \label{eq:M(z)}
\frac{M(r)}{M_0}=1+\delta_1 G(r;a,b;\ldots),
\end{equation}
where $\delta_1$ is a constant parameter, and $G(r;a,b;\ldots)$  encapsulates all the information on the geometry.

\section{The Born-Infeld model} \label{sec:VI}

The Born-Infeld (BI) Lagrangian \cite{BI34} is given, in its full form, by
\begin{equation} \label{eq:BI}
\varphi(X,Y)=2\beta^2 \left(1-\sqrt{1-\frac{X}{\beta^2}-\frac{Y^2}{4\beta^4}} \right),
\end{equation}
where $\beta$ is the BI parameter, such that (\ref{eq:BI}) recovers the Maxwell Lagrangian in the $\beta \rightarrow \infty$ limit. The BI Lagrangian arises in the low-energy regime of string and D-brane physics, where $\beta$ is related to the inverse string tension $\alpha'$ as $\beta=2\pi\alpha'$ \cite{string-NEDs1,string-NEDs2,string-NEDs3,string-NEDs4,string-NEDs5}. When coupled to gravity, the black holes associated to this Lagrangian have been studied in depth \cite{BI-GR1,BI-GR2,BI-GR3,BI-modifiedgravity1,BI-modifiedgravity2,BI-modifiedgravity3}.

In order to deal with dimensionless variables,  it is useful to define two length scales, $r_q^2\equiv \kappa^2q^2/4\pi$ and $l_\beta^2=(\kappa^2\beta^2/4\pi)^{-1}$, in terms of which we can define the dimensionless variable $z^4=\frac{r^4}{r_q^2l_\beta^2}$.  Using this, the electromagnetic field equation (\ref{eq:first-integral}) provides the expression
\begin{equation} \label{eq:field}
\frac{X}{\beta^2}=\frac{1}{1+z^4} \ ,
\end{equation}
which allows to express the components of the energy-momentum tensor associated to the Lagrangian (\ref{eq:BI}) as
\begin{equation} \label{eq:temBI}
\kappa^2T_t^t=\frac{1}{l_\beta^2} \left(1-\frac{\sqrt{z^4+1}}{z^2}\right) ; \kappa^2T_{\theta}^{\theta}=\frac{1}{l_\beta^2} \left(1-\frac{z^2}{\sqrt{z^4+1}} \right) \ .
\end{equation}
The trace (\ref{eq:trace}) can thus be written as
\begin{equation} \label{eq:traceBI}
\kappa^2T=-\frac{2}{l_\beta^2}\frac{\left(z^2-\sqrt{1+z^4}\right)^2}{z^2 \sqrt{1+z^4}},
\end{equation}
which is nonvanishing, as expected. Note in passing that in flat space, the energy density associated to a point-like charged field $F^{tr}(r)$ in this theory is obtained as
\begin{equation}\label{eq:nBI}
\varepsilon(q)= 4\pi \int_0^{\infty}dr r^2 T_t^t(r,q)=
\beta^{1/2}q^{3/2} \int_0^{\infty} dt (\sqrt{t^4+1}-t^2) =n_{BI}  \beta^{1/2} q^{3/2},
\end{equation}
where $n_{BI}=\frac{\pi^{3/2} }{3\Gamma(3/4)^2}\approx 1.23605$,  which yields a finite total energy.

\section{Quadratic gravity coupled to Born-Infeld} \label{sec:VII}

In this section we analyze the gravity theory (\ref{eq:gravitytheory}) coupled to Born-Infeld NED introduced in Sec. \ref{sec:VI}. This will allow us to present the basic equations of the problem and some useful transformations that will considerably simplify the algebraic expressions involved. Particular cases of interest will be considered in full detail later.

From the definitions introduced in Sec. \ref{sec:III}, the components of the metric $g_{\mu\nu}$ take the general form
\begin{eqnarray}
g_{tt}=-\frac{A(z)}{\sigma_{+}(z)} &;& g_{rr}=\frac{\sigma_{-}(z) }{\sigma_{+}(z)A(z)} \left(1+\frac{z\sigma_{-,z}}{2\sigma_{-}(z)} \right)^2\label{eq:BI1}\\
A(z)&=&1-\frac{1+\delta_1 G(z)}{\delta_2 z \sigma_{-}(z)^{1/2}} \label{eq:BI2} \ ,
\end{eqnarray}
where we have defined
\begin{equation}\label{eq:d1d2}
\delta_1=\frac{1}{2r_S}\sqrt{\frac{r_q^3}{l_{\beta}}} \ ; \ \delta_2= \frac{\sqrt{r_q l_{\beta}}}{r_S},
\end{equation}
and the function $G(z)$ satisfies
\begin{equation} \label{eq:Gz}
G_z=\frac{z^2 \sigma_{-}^{1/2}}{\sigma_+}\left(1+\frac{z\sigma_{-,z}}{2\sigma_-}\right)\left(l_{\beta}^2f+\tilde{\varphi}\right) \ .
\end{equation}
In this expression we have denoted  $\tilde{\varphi}=\varphi/\beta^2$.

In terms of the variable $z$, the expressions for $\sigma_\pm$ and $G_z$ for the theory (\ref{eq:gravitytheory}) for arbitrary values of the parameters $a$ and $b$ are very cumbersome, which complicates the identification of models that may contain geonic wormholes of the kind found in \cite{lor12}. For this reason, we find it very useful to perform a change of variable and introduce some redefinitions of the parameters of the theory. In particular, we consider the transformations
\begin{eqnarray}
z&=&\frac{\epsilon ^{1/2}}{\sqrt{2} (1+\epsilon )^{1/4}}  \label{eq:zepsilon}\\
\gamma_1&=& (2a+b)\lambda \\
\gamma_2&=& (4a+b)\lambda \ ,
\end{eqnarray}
where $\lambda\equiv l_P^2/l_\beta^2$.
With these definitions, we find that $G_\epsilon=G_z dz/d\epsilon$  reads as
\begin{equation}
G_\epsilon=\frac{\sqrt{\epsilon } \left(\epsilon  (2+\epsilon )^2+8 {\gamma_1} \right) \left((2+\epsilon )^3-8 {\gamma_1} (4+3 \epsilon ) \right)}{2 \sqrt{2} (1+\epsilon )^{7/4} (2+\epsilon )^2 (\epsilon  (2+\epsilon )+8 {\gamma_1})^2}\frac{\sigma_+}{\sigma_-^{1/2}},
\end{equation}
whereas $\sigma_\pm$ become
\begin{equation}
\sigma_+= \frac{\epsilon  (2+\epsilon ) (16{\gamma_1} +\epsilon  (2+\epsilon +8 {\gamma_1}  -4 \gamma_2 ))+32 {\gamma_1} \gamma_2  }{\epsilon  (2+\epsilon ) \left(2 \epsilon +\epsilon ^2+8 {\gamma_1}  \right)},
\end{equation}
and
\begin{equation}
\sigma_-= \frac{\epsilon  (2+\epsilon ) (8 {\gamma_2}  +\epsilon  (2+\epsilon -8 {\gamma_1}  +4 {\gamma_2} ))+32 {\gamma_1} {\gamma_2}  }{\epsilon  (2+\epsilon ) \left(2 \epsilon +\epsilon ^2+8 {\gamma_1} \right)}.
\end{equation}
Note that  when $\epsilon\gg 1$ we have $\sigma_\pm\to 1$ and $G_\epsilon d\epsilon/dz \approx 2/\sqrt{\epsilon}$, which nicely recovers the GR limit when (\ref{eq:zepsilon}) is used.

\section{Exactly solvable model} \label{sec:VIII}

To proceed further, it is convenient to specify particular values for the parameters $a$ and $b$ of the gravity Lagrangian (\ref{eq:gravitytheory}). For simplicity, we consider the case $f(R,Q)=R-\frac{l_P^2}{2}R^2+l_P^2 Q$, which provides useful simplifications that allow to make significant progress using analytical methods. This model corresponds to $a=-b/2$ and $b=1$ (or, equivalently,  $\gamma_1=0$ and $\gamma_2=-\lambda$, where $\lambda>0$). Any other choice of $b>0$ would be physically equivalent to this choice, up to a rescaling of the parameter $\lambda$. It should be noted that the choice $b>0$ was crucial in \cite{Barragan2010,Barragan2010b} to obtain bouncing cosmological models without big bang singularity. This motivates our choice\footnote{We note that exact solutions with $b<0$ can also be found, though they lead to physically uninteresting situations.}.
 For our choice of parameters, the general expressions given above boil down to
\begin{eqnarray}\label{eq:gzab}
G_\epsilon&=& \frac{2+\epsilon +\epsilon _c}{2 \sqrt{2} (1+\epsilon )^{7/4} \sqrt{\epsilon -\epsilon _c}}\\
\sigma_+&=& 1+\frac{\epsilon _c}{2+\epsilon } \\
\sigma_-&=& 1-\frac{\epsilon _c}{\epsilon } \ ,
\end{eqnarray}
where $\epsilon _c=4\lambda$ represents the point where $\sigma_-$ vanishes.
Similarly as in the Maxwell case \cite{lor12}, the existence of a zero in $\sigma_-$ implies that the region $\epsilon<\epsilon_c$ is not physically accessible\footnote{We note that in a spherically symmetric space-time, the area of the two spheres is given by $A=4\pi r^2(t,x)$, where $(t,x)$ are generic coordinates of the non-spherical sector. In static situations, we have $r^2(t,x)\to r^2(x)$ and one might want to use the function $r(x)$ as a coordinate of the non-spherical sector (canonical coordinates \cite{Stephani}). This is possible as long as $dr/dx\neq 0$. If in some interval $dr/dx=0$ at some point, then the function $r$ cannot be used as a coordinate on the whole interval. We will see later that in our problem the function $r(x)$ reaches a minimum ($dr/dx=0$ and $d^2r/dx^2>0$) and cannot cover the region $r<r_c$ (see Fig.\ref{fig:x(z)}). This explains why $\epsilon$ cannot be extended below $\epsilon_c$. The existence of a minimum area signals the existence of a minimum volume in the theory, an aspect related with the properties of the connection \cite{or13}. }, i.e., the geometry is only defined for $\epsilon\ge\epsilon_c$. In terms of $z$, the point $\epsilon_c$ corresponds to $z_c=\frac{ \sqrt{2\lambda }}{(1+4 \lambda )^{1/4}}$. Recall that the GR limit is recovered when $\epsilon\gg \epsilon _c$. For completeness, we note that for this model $G_z(\lambda)$ takes the form
\begin{equation} \label{eq:Gz-general}
G_z=\frac{2z W[z]\left( z^2+(1+2 \lambda )W[z]\text{  }\right)}{\sqrt{1+z^4}\sqrt{z^2-2 \lambda  W[z]}} \ ,
\end{equation}
where $W[z]\equiv \sqrt{1+z^4}-z^2$. From this expression, one readily verifies that the GR limit is correctly recovered in the limit  $\lambda \rightarrow 0$:
\begin{equation} \label{eq:Gz-GR}
G_z\approx 2 \left(\sqrt{1+z^4}-z^2\right)+ O(\lambda) \ .
\end{equation}
The integration of (\ref{eq:gzab}) is immediate and yields
\begin{equation}
G(\epsilon)=-\frac{1}{\delta_c}+\frac{\sqrt{2(\epsilon -\epsilon _c)} }{3 (1+\epsilon )^{3/4}}\Big[1+ {_{2}{F_1}}\left[\frac{1}{2},\frac{3}{4},\frac{3}{2},-\frac{ (\epsilon -\epsilon_c)}{1+\epsilon _c}\right]2 \left(\frac{1+\epsilon }{1+\epsilon _c}\right){}^{3/4}\Big],
\end{equation}
where ${_{2}{F_1}}$ represents a hypergeometric function, and $\delta_c$ is an integration constant. In order to recover the correct behavior at infinity, $G(z)\approx -1/z$, one finds that
\begin{equation}\label{eq:dBI}
\delta_c=\frac{(1+4\lambda)^{1/4}}{2 n_{BI}} \ ,
\end{equation}
where $n_{BI}$ was defined in (\ref{eq:nBI}).

\begin{figure}[tbp]
\centering
\includegraphics[width=0.5\textwidth]{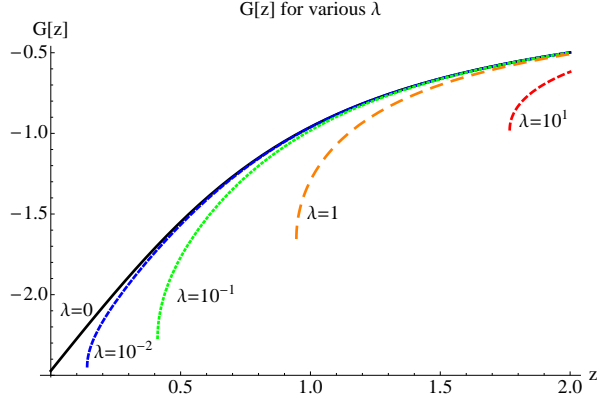}
\caption{Evolution of the function $G(z)$ for different values of $\lambda$. The GR limit corresponds to $\lambda=0$. For a given $\lambda$, the minimum occurs at $z_c=\frac{ \sqrt{2\lambda }}{(1+4 \lambda )^{1/4}}$ and corresponds to $G(z_c)=-1/\delta_c$ [see (\ref{eq:dBI})]. Note that in GR the geometry extends down to $z=0$, whereas for $\lambda>0$ we find that $z$ cannot be smaller than $z_c$, which manifests the existence of a finite structure (or core) replacing the central point-like singularity. \label{fig:G}}
\end{figure}

\subsection{Metric components. \label{sec:metric_expansions}}

When $\gamma_1=0$, one finds that $g_{rr}=-1/(g_{tt} \sigma_-)$. It is thus useful to define a new radial coordinate such that $g_{rr}dr^2=-dx^2/g_{tt}$, which brings the line element into its usual Schwarzschild-like form
\begin{equation}\label{eq:ds2x}
ds^2=g_{tt}dt^2-\frac{1}{g_{tt}}dx^2+r^2(x)d\Omega^2 \ .
\end{equation}
 All the information about the geometry is thus contained in the functions $g_{tt}=-{A}/{\sigma_{+}}$ and $r^2(x)$. In our case, the relation between $x$ and $r$ can be explicitly written in terms of a hypergeometric function whose asymptotic behaviors are $x\approx r$ for large values of $r$, and $x/r_c\approx x_0(\lambda)+\alpha_\lambda \sqrt{z-z_c}+\ldots$, where $x_0(\lambda)\approx-1.694 \lambda (1 + 4 \lambda)^{-3/4}$ and $\alpha_\lambda>0$ as $z\to z_c$.

From the previous results, one can obtain series expansions that allow to study the asymptotic behavior of the geometry when $z\to \infty$, and also in the vicinity of the core, namely, $z\to z_c$. In the far limit, we find
\begin{equation}
g_{tt}\approx  -1  +\frac{1}{\delta _2 z}   -\frac{\delta _1}{\delta _2 z^2}  +  \frac{\lambda }{z^4}+ O\left(\frac{1}{z^5} \right)  \ .
\end{equation}
This expression exactly recovers the result of GR coupled to the BI theory. In the region near the core, the metric takes the form
\begin{equation}
g_{tt}\approx \frac{(1+2 \lambda )^{3/2} \left(\delta _c-\delta _1\right)}{2^{3/4} \lambda ^{1/4} (1+4 \lambda )^{11/8} \delta _2 \delta _c \sqrt{z-z_c}}+ \frac{(1+2 \lambda ) \left(2 \delta _1-\sqrt{1+4 \lambda } \delta _2\right)}{(1+4 \lambda )^{3/2} \delta _2} + O(\sqrt{z-z_c})\ .
\end{equation}
From this expression we see that, in general, the metric diverges at $z_c$ as $\sim 1/\sqrt{z-z_c}$. However, for the choice $\delta _1=\delta _c$, one finds a finite result everywhere
\begin{equation} \label{eq:metricregular}
g_{tt}\approx  \frac{(1+2 \lambda ) \left(-\sqrt{1+4 \lambda } \delta _2+2 \delta _c\right)}{(1+4 \lambda )^{3/2} \delta _2}+O(z-z_c) \ .
\end{equation}
Note that the choice  $\delta_1=\delta_c$ allows to turn the metric near $z=z_c$ into Minkowskian form by just rescaling by a constant factor the $t$ and $x$ coordinates. To better understand the details of the geometry in this region, we proceed next to evaluate the curvature invariants of the metric.

\subsection{Curvature invariants \label{sec:curvature_expansions}}

To characterize the geometry in a coordinate-independent manner, we consider the scalars  $R(g)\equiv g^{\mu\nu}R_{\mu\nu}(g)$,  $Q(g)\equiv R_{\mu\nu}(g)R^{\mu\nu}(g)$, and $K(g)\equiv R_{\alpha\beta\gamma\delta}(g)R^{\alpha\beta\gamma\delta}(g)$ constructed out of the physical metric $g_{\mu\nu}$. Expansion of these scalars in the far region ($z\gg z_c$) leads to
\begin{eqnarray}
R(g)&=& \frac{r_q^4 l_\beta^2}{2r^8}\left(1-\frac{28  l_P^2}{r^{2}}+\ldots\right)  \\
Q(g)&=& \frac{r_q^4}{r^8}\left(1-\frac{16  l_P^2}{r^{2}}+\ldots\right) \\
K(g)&=& K_{GR}+\frac{144 r_S r_q^2 l_P^2}{r^9}+\ldots   \ ,
\end{eqnarray}
where $K_{GR}$ represents the GR value of the Kretschmann scalar. It is worth noting that in $Q(g)$ and in $K(g)$ the Planck scale corrections decay at a slower rate than those due to the Born-Infeld contribution, whereas in $R(g)$ the Planckian contribution decays much faster.

On the other hand,  in the region $z\approx z_c$ we find that, in general, the expansions of these invariants around $z=z_c$ can be written as
\begin{eqnarray}
R&=& (\delta_1-\delta_c)\left[\frac{\alpha_1}{(z-z_c)^{\frac{3}{2}}} + \frac{\alpha_2}{(z-z_c)^{\frac{1}{2}}}+O[(z-z_c)^{\frac{1}{2}}]\right] + C_1+\ldots \\\
Q&=&   (\delta_1-\delta_c)^2\left[\frac{\zeta_1}{(z-z_c)^3} +\frac{\zeta_2}{(z-z_c)^2}+\frac{\zeta_3}{(z-z_c)^1}\right]+\nonumber\\&+&  (\delta_1-\delta_c)\left[\frac{\xi_1}{(z-z_c)^{\frac{3}{2}}} +\frac{\xi_2}{(z-z_c)^{\frac{1}{2}}} +O[(z-z_c)^{\frac{1}{2}}]\right] + C_2+\ldots   \\
K&=&   (\delta_1-\delta_c)^2\left[\frac{\mu_1}{(z-z_c)^3} +\frac{\mu_2}{(z-z_c)^2}+\frac{\mu_3}{(z-z_c)^1}\right]+\nonumber\\&+&  (\delta_1-\delta_c)\left[\frac{\nu_1}{(z-z_c)^{\frac{3}{2}}} +\frac{\nu_2}{(z-z_c)^{\frac{1}{2}}} +O[(z-z_c)^{\frac{1}{2}}]\right] +  C_3 + \ldots  \ .
\end{eqnarray}
where $\alpha_i, \zeta_i$, $\xi_i$, $\mu_i$, $\nu_i$, and $C_i$ are constants that depend on $(\lambda,\delta_1,\delta_2,\delta_c)$. We thus see that these three curvature invariants are divergent at $z=z_c$, with the leading terms growing as $\sim 1/(z-z_c)^3$ in the worst case. This behavior is similar to that found in \cite{lor12} for the case of Maxwell electrodynamics and contrasts with the results of GR, where the divergence grows as $1/z^8$ in the case of Maxwell, and as $1/z^4$ in the case of Born-Infeld. Of particular interest is the case  $\delta_1=\delta_c$, for which the expansions of the curvature invariants around $z=z_c$ read
\begin{eqnarray}
R&\approx& \frac{4 \left(16 \lambda ^2+11 \lambda +3\right) \delta _c-3 \delta _2 \sqrt{4 \lambda +1} \left(8 \lambda ^2+4 \lambda +1\right)}{3  \lambda  (2 \lambda +1)^2\delta _2} + O(z-z_c) \\
Q&\approx&D (9  \left(160 \lambda ^5+232 \lambda ^4+160 \lambda ^3+60 \lambda ^2+12 \lambda +1\right) \delta _2^2 \nonumber \\ &-&  12  \sqrt{4 \lambda +1} \left(104 \lambda ^4+128 \lambda ^3+78 \lambda ^2+23 \lambda +3\right) \delta _c \delta _2  \nonumber \\ &+&  \left.2 \left(688 \lambda ^4+976 \lambda ^3+652 \lambda ^2+204 \lambda +27\right) \delta _c^2 \right) \\ &+& O(z-z_c) \nonumber \\
K&\approx& \frac{4\lambda+1}{\lambda^2}+ \frac{2 \left(\delta _2 \sqrt{4 \lambda +1}-2 \delta _c\right)^2}{\lambda ^2\delta _2^2} + \frac{16 \left((1-4 \lambda ) \delta _c+3 \delta _2 \lambda  \sqrt{4 \lambda +1}\right)^2}{9  (2 \lambda +1)^4\delta _2^2} \nonumber \\ &+&  O(z-z_c),
\end{eqnarray}
where $D= \frac{1}{9 \delta _2^2 \lambda ^2 (2 \lambda +1)^4}$. This confirms that, like in the Maxwell case \cite{lor12}, there exist configurations for which the geometry at $z=z_c$ is completely regular.

\subsection{Wormhole structure and electric flux}

\begin{figure}[tbp]
\centering
\includegraphics[width=0.5\textwidth]{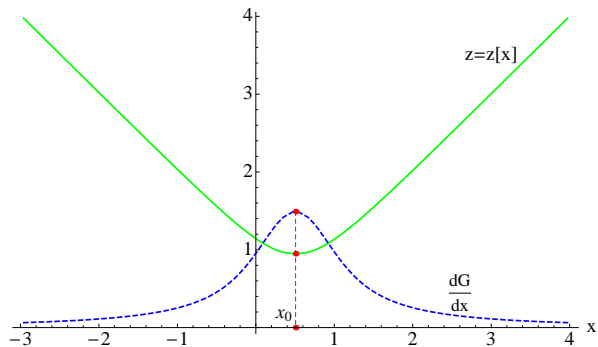}
\caption{Representation of the radial function $z$ in terms of $x$. The minimum of the curve represents the wormhole throat, where $dz/dx=\sigma_-^{1/2}=0$ and $d^2z/dx^2>0$. Note that $G_x$ is smooth and finite everywhere. \label{fig:x(z)}}
\end{figure}

The existence of solutions with a smooth geometry at $z=z_c$ indicates that one can still use the field equations of the theory to investigate the physical processes going on in that region. In particular, one may be interested in the nature of the sources that generate the electric field. In this sense, one should note that we are dealing with sourceless electromagnetic field equations. This fact, together with the constraints imposed by the field equations that force $z$ to be greater or equal than $z_c$, suggests that the spherically symmetric electric field could have a topological origin rather than being created by a distribution of charges, an idea dating back to Misner and Wheeler \cite{Misner-Wheeler1957}, and that our model implements in a natural way. In fact, the absence of singularities at $z=z_c$ drive us to consider an extension of the geometry allowed by the definition of the coordinate $x$ introduced in (\ref{eq:ds2x}) such that $x$ can be extended to the whole real axis while $z$ remains bounded to the region $z\ge z_c$. This can be done by considering the relation $dx^2=dz^2/\sigma_-$ and noting that two signs are possible in the resulting definition of $x(r)$, namely,  $dx=\pm dz/\sigma_-^{1/2}$. In our analysis of the curvature invariants and the metric above, we assumed $dx=+ dz/\sigma_-^{1/2}$. The existence of nonsingular solutions motivates the consideration of $dx=- dz/\sigma_-^{1/2}$ as a physically meaningful branch of the theory. This leads to (see Fig.\ref{fig:x(z)})
\begin{equation}
x(z)=\left\{\begin{array}{lr} F(z;\lambda) & \text{ if } x\ge  x_c \\
                                              2x_c -F(z;\lambda)  & \text{ if } x\le x_c\end{array}\right.
\end{equation}
where $F(z_c;\lambda)=x_c=\sqrt{\frac{\pi }{8}}\frac{\Gamma \left[-\frac{1}{4}\right]}{ \Gamma \left[\frac{1}{4}\right]}\frac{ z_c^2  }{(1+4 \lambda )^{1/4}}$, and
\begin{eqnarray}
F(z;\lambda)&=&x_c+\frac{\sqrt{\epsilon -\epsilon _c} }{2 \sqrt{2} (1+\epsilon )^{1/4} \left(1+\epsilon _c\right)}\Big(2+_2F_1\left[\frac{1}{4},\frac{1}{2},\frac{3}{2},-\frac{\epsilon -\epsilon _c}{1+\epsilon _c}\right] \epsilon _c \left(\frac{1+\epsilon }{1+\epsilon _c}\right){}^{1/4}\Big) \ . \nonumber
\end{eqnarray}

Therefore, to cover the whole geometry, one needs two charts if the set $(t,r)$ is used as coordinates (one for the interval $x\ge  x_c$ and another for $x\le  x_c$) or a single chart if $(t,x)$ is used instead.

The bounce of the radial function $r^2(x)$ (see Fig.\ref{fig:x(z)}) puts forward that our spacetime has a genuine wormhole geometry (see \cite{Lobo} for a review of wormhole solutions in the literature). As a result, the spherically symmetric electric field that local observers measure is not generated by a charge distribution but by a sourceless electric flux trapped in the topology \cite{Misner-Wheeler1957}. In this sense, an observer in the $x>x_c$ region finds that the electric flux across any closed 2-surface containing the wormhole throat in its interior is given by $\Phi\equiv \int_S \varphi_X  *F=4\pi q$, where $*F$ represents the 2-form dual to Faraday's tensor. If the integration is performed in the $x<x_c$ region and the orientation is assumed such that the normal points in the direction of growth of the area of the 2-spheres,  then the result is $\Phi=-4\pi q$ because the orientation differs in sign with that chosen on the other side of the throat.  This shows that  no {\it real} sources generate the field, which is fully consistent with the sourceless gravitational-electromagnetic equations of our theory.

We note that the electric flux per surface unit flowing through the wormhole throat at $z=z_c$ takes the form
\begin{equation}\label{eq:dens_flux}
\frac{\Phi}{4\pi r_c^2 z_c^2}=\left(\frac{4\lambda}{1+4\lambda}\right)^{1/2}\sqrt{\frac{c^7}{2(\hbar G)^2}} \ .
\end{equation}
This quantity represents the density of lines of force at $z_c$. For any fixed $\lambda$, it turns out to be a universal constant that only depends on $\hbar$, $c$, and $G$. When $\lambda\to \infty$, which is equivalent to taking the limit in which the Born-Infeld theory recovers Maxwell's electrodynamics ($l_\beta\to 0$), we reproduce the result obtained in \cite{lor12}. If one considers the limit $\lambda\to 0$, which corresponds to the coupling of the Born-Infeld theory to GR ($l_P\to 0$), one finds that the density of lines of force tends to zero. This situation can be seen as the limit in which the wormhole disappears ($z_c\to 0$), which is consistent with studies showing the impossibility of generating wormholes supported by nonlinear theories of electrodynamics in GR \cite{Arellano1,Arellano2}. As a result,
in the case of GR (coupled to Maxwell or to BI), one needs to find a source for the existing electric fields. This leads to a well-known problem plagued by inconsistencies related to the impossibility of having a point-like particle at rest at $r=0$ whose energy and charge match those appearing in the geometry and, at the same time, being a solution of the Einstein field equations everywhere. The wormhole structure that arises in our theory, therefore, naturally avoids the problem of the sources that one finds in GR \cite{Ortin}.

The fact that for any $\lambda>0$ the density of lines of force at the wormhole throat is a constant independent of the particular amounts of charge and mass strongly supports the view that the wormhole structure exists even in those cases in which $\delta_1\neq \delta_c$. To the light of this result, the question of the meaning and implications of curvature divergences should be reconsidered, since their presence seems to pose no obstacle to the existence of a well-defined electric flux through the $z=z_c$ surface.

\subsection{Horizons and mass spectrum when  $\delta_1= \delta_c$}

From the analysis in sections \ref{sec:metric_expansions} and \ref{sec:curvature_expansions}, one can easily verify that, regardless of the value of $\delta_1$,  a few $z_c$ units away from the center the curvature invariants  and the location of the external event horizon rapidly tend to those predicted by GR.  However, for solutions with $\delta_1= \delta_c$ one finds that the event horizon may disappear in certain cases, which brings about a new kind of  gravitating object whose internal and external properties differ from those typically found in GR.  In fact, one can verify numerically that for such configurations the existence of the event horizon crucially depends on the sign of the leading term of $g_{tt}$ in (\ref{eq:metricregular}). If this term is positive, then there exists an external horizon, but if it is negative then the horizon is absent. This implies that the horizon disappears if the condition
\begin{equation}\label{eq:ineq0}
2\delta_c<\delta_2\sqrt{1+4\lambda}
\end{equation}
is satisfied. This condition can be put in a more intuitive form if
one expresses the charge as $q=N_q e$, with $N_q$ representing the number of elementary charges $e$, which leads to $r_q=2l_P N_q/N_q^M$, where $N_q^M\equiv \sqrt{2/\alpha_{em}}\approx 16.55$ and $\alpha_{em}$ is the fine structure constant. Taking now into account the definitions  given in (\ref{eq:d1d2}), one finds that $\delta_1/\delta_2=\lambda^{1/2} N_q/N_q^M$, which allows to rewrite  (\ref{eq:ineq0}) as
\begin{equation}
N_q< N_q^M\sqrt{1+\frac{1}{4\lambda}} \ .
\end{equation}
Therefore, if the number of charges of the object is smaller than  $N_q^{BI}=N_q^M\sqrt{1+\frac{1}{4\lambda}}$ the object has no external event horizon and appears to an external observer in much the same way as a massive, charged elementary particle.

The mass spectrum of these objects can be obtained directly from the regularity condition $\delta_1=\delta_c$, which establishes a constraint between the amount of charge and mass of those solutions. Using again the notation $q=N_q e$, we find
\begin{equation}\label{eq:M}
M^{BI}=\left(1+\frac{1}{4\lambda} \right)^{1/2} \left(\frac{N_q}{N_q^{BI}}\right)^{3/2} n_{BI} m_P,
\end{equation}
where $n_{BI}\approx 1.23605$ was defined in (\ref{eq:nBI}). This
mass spectrum can also be written as
\begin{equation}\label{eq:M2}
M^{BI}=\left(\frac{4\lambda}{1+4\lambda}\right)^{1/4} M^M \ ,
\end{equation}
where $M^M=(N_q/N_q^M)^{3/2} n_{BI}m_P$ is the mass spectrum of the Maxwell case found in \cite{lor12}. Note that the limit $\lambda\to \infty$ nicely recovers the result of the Maxwell case (recall that $\lambda\equiv l_P^2/l_\beta^2$).  In general, we see that for any finite $\lambda$, the mass corresponding to a given charge $q$ is smaller in the BI than in the Maxwell case. This manifests that the quantum effects responsible for the nonlinearities of the matter field have an impact on its energy density and  result in a spectrum of lighter particles.

To illustrate this last point, let us estimate some typical scale for the parameter $\beta$. This can be done from the effective Lagrangians scheme of quantum electrodynamics \cite{Dobado}, which is a reliable approximation as far as the maximum allowed electric field does not exceed the range $\sim 10^{16}-10^{18} $, and that allows to employ (nonlinear) classical Lagrangians as suitable phenomenological descriptions of the electromagnetic field, encoding quantum vacuum effects. Noting that $\beta$ is identified as the maximum value of the electric field in BI theory, attained at the center [see Eq.(\ref{eq:field})] this sets a maximum scale for $\beta$ in the range above (note also that the expansion of BI Lagrangian for small $\beta^2$ coincides, for $Y=0$, and to lowest order in the perturbative expansion \cite{Dobado}, with the Euler-Heisenberg effective Lagrangian of quantum electrodynamics \cite{EH1,EH2}, modulo a constant). Since $\beta^2$ has dimensions of $q^2/c r^4$, where $c$ is the speed of light, we can see the value of $\beta^2$ as the intensity of electric field (squared) that an electron generates at a distance $r_e$ such that $r_e^4=e^2/c \beta^2=\alpha_{em} \hbar/\beta^2$. From the definition of $\lambda\equiv l_P^2/l_\beta^2$ and $l_\beta^2=4\pi/(\kappa^2 \beta^2)=c^3/(2G \beta^2)$, we find that $l_\beta^2=c^3 r_e^4/(2G \alpha_{em} \hbar)=r_e^4/2 l_P^2$. With this result, we can write $\lambda=2 l_P^4/r_e^4$. For $\beta=10^{18}$, we get $r_e\approx 10^{-18}$ m, which leads to $\lambda\sim (10^{-17})^4$. The smallness of this parameter implies that $M^{BI}\approx (4\lambda)^{1/4}M^M\approx 10^{-17} M^M$, i.e., the mass has been lowered by $17$ orders of magnitude! This brings the mass spectrum of our solutions from the Planck scale, $10^{19}$ GeV, down to the reach of current particle accelerators, $\sim 10^{2}$ GeV, which shows  that the Planck scale phenomenology of Palatini gravity can be tested and constrained with currently available experiments.

We note that the matter model used in our discussion is far from being completely satisfactory and that a number of  corrections might be important in the range of scales considered.  In fact, more accurate descriptions of the quantum vacuum effects would require taking into account strong field and nonperturbative effects, aspects that are still under intense scrutiny (see, for instance, \cite{Bastianelli1,Bastianelli2,Bastianelli3}). In this sense, the BI toy model discussed here must be seen as an approximation which by no means incorporates all the effects expected to be relevant in this problem. Nonetheless, it puts forward the important phenomenological effects that non-linearities in the matter sector could have if at relatively low energies the quantum gravitational degrees of freedom required a metric-affine structure to accurately describe the space-time dynamics.

\subsection{Geons as point-like particles}

The regular solutions that we are considering represent an explicit realization of the concept of geon introduced by Wheeler in \cite{Wheeler1955}. These objects were defined as self-gravitating, nonsingular solutions of the sourceless gravitational-electromagnetic field equations. In our case, however, the regularity condition seems not to be so important to obtain consistent solutions to the gravitational-electromagnetic system of equations since, as we have shown, the existence of curvature divergences at the wormhole throat has no effect on the properties of the electric field that flows through it. Nonetheless, the existence of completely regular horizonless solutions that could be interpreted as elementary particles with mass and charge suggests that such solutions could play a more fundamental role in the theory. Further evidence in this direction can be obtained by evaluating the full action on the solutions that we have found. This computation will give us information about the total electric and gravitational energy stored in the space-time. \\
With elementary manipulations, we find that
\begin{equation}
S=\frac{1}{2\kappa^2} \int d^4x \sqrt{-g}(R-\frac{l_P^2}{2}R^2+l_P^2 Q) +  \frac{1}{8\pi}\int d^4 x \sqrt{-g}  \varphi = - \int d^4 x \sqrt{-g} T_t^t \sigma_{+} \ ,
\end{equation}
where we have replaced $R$ and $Q$ by their respective expressions (\ref{eq:curvaturescalar}) and (\ref{eq:Q-invariant}), and have also used that $\varphi/8\pi=T_{\theta}^{\theta}$. Carrying out explicitly the integration over the whole spacetime in terms of the variable $dx^2=dz^2/\sigma_{-}$  leads to
\begin{eqnarray} \label{eq:actionderivation}
S&=& - (r_ql_{\beta})^{3/2}  \int dt \int _{-\infty}^{+\infty} dx z^2 d\Omega T_t^t \sigma_{+} = -8\pi (r_ql_{\beta})^{3/2} \int dt \int_{z_c}^{+\infty} \frac{dz z^2}{\sigma_{-}^{1/2}} T_t^t \sigma_{+} \\
&=& 4Mc^2 \delta_1 I \int dt, \nonumber
\end{eqnarray}
where $I=-\kappa^2 l_{\beta}^2 \int_{z_c}^{+\infty} dz z^2 T_t^t \sigma_{+}/\sigma_{-}^{1/2}$ only depends on $\lambda$. In the second step in (\ref{eq:actionderivation}) we have introduced a factor $2$ in order to take into account the two sides of the wormhole, and in the last step we have used the definitions of Eqs.(\ref{eq:d1d2}) and (\ref{eq:temBI}). Note that $M$ represents the integration constant that we identified with the Schwarzschild mass of the uncharged solution. Making use of the variable $\epsilon$ and applying the theorem of residues, we find $I=n_{BI}/(1+4\lambda)^{1/4}=1/(2\delta_c)$ [see Eq.(\ref{eq:dBI})] and thus we collect the final result
\begin{equation}
S=2Mc^2 \frac{\delta_1}{\delta_c} \int dt.
\end{equation}
This implies that when the regularity condition $\delta_1=\delta_c$ holds the resulting action reads
\begin{equation}
S=2M^{BI}c^2  \int dt \ ,
\end{equation}
with $M^{BI}$ being the mass defined in (\ref{eq:M}).
Remarkably, this is just the action of a point-like massive particle at rest. Taking into account that local observers can only be on one of the sides of the wormhole,
it follows that one factor $M^{BI}$ comes from integrating on one side of the space-time and the other comes from integrating on the other side. In this sense, the spatial integral
\begin{equation}
\E= \int dx z^2 d\Omega\left[ \frac{1}{2\kappa^2}f(R,Q) +  \frac{ \varphi }{8\pi} \right]
\end{equation}
can be seen as the addition of the electromagnetic energy plus the gravitational binding energy generated by the electric field. Since in the regular cases, $\delta_1=\delta_c$, the total energy produced by the electric field is equal to the Schwarzschild mass of the object, $M=M^{BI}$, it follows that these configurations are, in fact, geonic-like gravitational solitons. This result further supports our view that these objects, with or without an event horizon, have particle-like properties. Additional evidence relevant for this discussion comes from the fact that the hypersurface $x=x_c$, where the wormhole throat is located, changes from space-like to time-like when the horizon disappears. This means that the wormhole throat in the horizonless solutions follows a time-like trajectory, like any massive physical particle. Therefore, to an external observer with low resolution power, such objects would appear as massive, charged point-like particles.

It should be stressed that in the context of GR the identification between the total energy and the Schwarzschild mass is also obtained in some models where an exotic dust and a Maxwell field are considered [see e.g. \cite{Bronnikov})]. Similarly as in our case, the factor $2M$ comes from the integration of the radial coordinate along its whole range of definition, and can be interpreted as corresponding to a pair of particles, located on each side of the wormhole.

\section{Summary and conclusions} \label{sec:IX}

In this paper we have considered the coupling of a nonlinear theory of electrodynamics to a quadratic extension of GR formulated \`{a} la Palatini. Unlike in the standard formulation of quadratic gravity, in which fourth-order equations govern the behavior of the metric, in the Palatini formulation we find second-order equations for the metric. This follows from the fact that the connection can be solved by algebraic means as the Levi-Civita connection of an auxiliary metric $h_{\mu\nu}$ \cite{or13}. Since this metric satisfies second-order equations and is algebraically related with $g_{\mu\nu}$, the theory turns out to have the same number of propagating degrees of freedom as GR. Moreover, in vacuum the field equations of our theory exactly become those of GR, which is a manifestation of the observed universality of Einstein's equations in Palatini theories \cite{Ferraris:1992dx, Borowiec:1996kg}. The simplicity of the field equations governing $h_{\mu\nu}$ has allowed us to find exact formal solutions for arbitrary gravity Lagrangian $f(R,Q)$ and arbitrary electrodynamics theory $\varphi(X,Y=0)$ in the electrostatic case.

The introduction of a nonlinear electrodynamics theory is motivated by the need to consider quantum corrections in the matter sector at energies above the pair-production threshold. We were particularly interested in determining if the geon-like solutions found in a purely Maxwell context in the quadratic Palatini theory would persist under quantum-induced modifications of the matter sector. We have found that this is indeed the case. Geon-like solutions consisting on a wormhole supported by a Born-Infeld electrostatic field exist in a family of gravity models whose parameters also yield nonsingular, bouncing cosmologies ($b>0$) \cite{Barragan2010}. For simplicity, we have considered a specific choice of parameters $(a=-b/2, b=1)$, which yields exact analytical solutions and allows to obtain definite predictions about the charge and mass spectrum of the theory.

In this respect, we have found that the mass spectrum of the regular geons, the ones without curvature divergences at the wormhole throat, can be several orders of magnitude smaller than those found in Maxwell's theory. For values of the parameters reaching the limit of validity of the approximations where the use of classical nonlinear Lagrangians such as the Born-Infeld model is justified, we have shown that the mass spectrum can be lowered from the Planck scale down to the GeV scale. There is a simple physical reason for this effect. For small values of $\lambda$ the density of lines of force at the wormhole throat [see Eq. (\ref{eq:dens_flux})] decreases. Since, as we have shown, the mass of these objects is due to the energy stored in the electrostatic field,  lowering the density of lines of force implies lowering their mass. In the GR limit, $\lambda\to 0$, the wormhole closes and these objects disappear from the spectrum of solutions of the theory.

In our view, the particular combination of theories considered here must be regarded as a toy model that could be improved in different ways. Nonetheless, it allows to put forward that new avenues to the generation of stable massive particles from a quantum gravitational perspective are possible and that  Planck scale physics can be brought into experimental reach within a purely four-dimensional scenario. The potential implications that the existence of stable massive particles, such as the horizonless geons found here, could have for the missing matter problem in astrophysics and cosmology \cite{Astro1,Astro2} demand further research in this direction. In particular, improvements in the treatment of the quantum corrections affecting the matter sector and a rigorous derivation of the semiclassical corrections expected in Palatini backgrounds should be considered in detail in the future.

To conclude, we underline that the results of this paper indicate that the existence of geons is not directly tied to the particular matter source considered, because such solutions do not arise in GR \cite{Arellano1,Arellano2}, but rather to the Planck-corrected Palatini model employed. The matter is needed to excite the new gravitational dynamics (recall that in vacuum the theory boils down to GR)  but the existence of wormholes to support the geons must be attributed to the quantum-corrected gravity Lagrangian and the way Palatini dynamics affects the structure of space-time. This suggests that similar phenomenology could be expected when non-abelian gauge fields are added to the matter sector and also in other approaches based on the premise that metric and affine structures are {\it a priori} independent \cite{PhenomQG}.

\section*{Acknowledgments}

G. J. O. is finantially supported by the Spanish grant FIS2011-29813-C02-02  and the JAE-doc program of the Spanish Research Council (CSIC).  D. R. -G. is supported by CNPq (Brazilian agency) through grant number 561069/2010-7 and acknowledges the hospitality and partial support of the theoretical physics group at the University of Valencia, where this work was initiated. We are indebted to E. Guendelman, and M. Vasihoun for useful discussions and comments, and to F. S. N. Lobo for his critical reading of the manuscript.

\end{document}